\documentclass[10pt,sigconf]{acmart}

\usepackage{subcaption}
\usepackage{booktabs}
\usepackage{multirow}
\usepackage{xcolor}
\usepackage[ruled,vlined,linesnumbered]{algorithm2e}
\usepackage{pifont}
\newtheorem{theorem}{Theorem}

\begin{document}

\title{Secure Transformer Inference Protocol}

\author{Mu Yuan}
\email{ym0813@mail.ustc.edu.cn}
\affiliation{%
    \institution{University of Science and Technology of China}
    \city{Hefei}
    \country{China}
}
\author{Lan Zhang}
\email{zhanglan@ustc.edu.cn}
\affiliation{%
  \institution{University of Science and Technology of China}
  \city{Hefei}
  \country{China}
}
\author{Xiang-Yang Li}
\email{xiangyangli@ustc.edu.cn}
\affiliation{%
    \institution{University of Science and Technology of China}
    \city{Hefei}
    \country{China}
}

\newcommand{\myhighlight}[1]{
    \\\noindent\rule{0.99\linewidth}{0.5pt}\\
    \textit{\textbf{$\triangleright$ #1}}\\
    \noindent\rule{0.99\linewidth}{0.5pt}\\
}

\newcommand{\mysys}{STIP~}

\definecolor{ao}{rgb}{0.0, 0.5, 0.0}
\definecolor{britishracinggreen}{rgb}{0.0, 0.26, 0.15}
\definecolor{carnelian}{rgb}{0.7, 0.11, 0.11}

\newcommand{\cmark}{{\color{ao} \ding{51}}}%
\newcommand{\xmark}{{\color{carnelian} \ding{55}}}%

\begin{abstract}
Security of model parameters and user data is critical for Transformer-based services, such as ChatGPT.
While recent strides in secure two-party protocols have successfully addressed security concerns in serving Transformer models, their adoption is practically infeasible due to the prohibitive cryptographic overheads involved.
Drawing insights from our hands-on experience in developing two real-world Transformer-based services, we identify the inherent efficiency bottleneck in the two-party assumption. 
To overcome this limitation, we propose a novel three-party threat model. 
Within this framework, we design a semi-symmetric permutation-based protection scheme and present STIP, the first secure Transformer inference protocol without any inference accuracy loss.
Experiments on representative Transformer models in real systems show that STIP has practical security and outperforms state-of-the-art secure two-party protocols in efficiency by millions of times.
\end{abstract}

\ccsdesc[500]{Networks~Network protocol design}
\ccsdesc[300]{Networks~Cloud computing}
\ccsdesc[500]{Networks~Security protocols}
\ccsdesc[500]{Security and privacy~Privacy-preserving protocols}

\maketitle

\section{Introduction}
\label{sec:intro}

Transformer inference-based services are the most cutting-edge artificial intelligence applications~\cite{chatgpt-record, gemini, bard}.
Cloud computing capabilities, such as auto-scaling~\cite{romero2021infaas}, meet the requirements of serving Transformers, especially large models with billions of parameters.
Therefore, major organizations like OpenAI opt for full-cloud deployment for their Transformer-based services~\cite{azure-openai}.
Nevertheless, sending raw data to the cloud is infeasible in privacy-sensitive domains, as illustrated by incidents such as Samsung's prohibition of ChatGPT use after a sensitive code leak~\cite{samsung}.

\textbf{Model split is not secure enough.}
Model split inference~\cite{dc-transformer, thapa2022splitfed, zeng2020coedge} strategically distributes neural network layers between the device and the cloud.
The device sends intermediate activations to the cloud to continue inference.
Model split inference avoids raw data transmission while maintaining efficiency~\cite{pham2023binarizing, pasquini2021unleashing}. 
However, concerns arise as research reveals the potential for reverse-engineering sensitive information from intermediate activations~\cite{abuadbba2020can, pan2020privacy}.

\textbf{Secure two-party protocols incur prohibitive overheads.}
Recent studies explore secure Transformer inference through homomorphic encryption (HE) and secure two-party computation (2PC)~\cite{hou2023ciphergpt, hao2022iron, chen2022x}.
However, these protocols incur prohibitive computational and device-cloud communication overheads, especially with non-linear complex layers like LayerNorm and ReLU. 
For example, CipherGPT costs a 25-minute processing time and 90 GiB traffic for generating a single token with a GPT2 model~\cite{hou2023ciphergpt}.

\textbf{First principles thinking: three-party threat model.}
To overcome the efficiency barrier, we use first principles to rethink the basic two-party assumption: model owner and data owner.
Drawing insights from managing two real-world Transformer-based services\footnote{Please refer to Sec.\ref{subsec:threep} for a detailed introduction of these two services.}, a consistent experience emerged: \textit{model developer $\neq$ model server}. 
For both services, we developed Transformer models by fine-tuning~\cite{lora} open-sourced parameters~\cite{touvron2023llama, liu2023llava} with collected data.
We have adequate computing power for offline model development but lack the capability for large-scale, long-term services to hundreds of thousands of users. 
Consequently, as model developers, we rely on third-party cloud platforms to serve our models. 
In line with our real-world development experiences, we propose a new three-party threat model.
Within this model, we decompose the model owner into two entities: model developer and model server. 
As the developed models are proprietary, model developers must safeguard their model parameters against potential attacks from model servers~\cite{model-confidential}, therefore we assume they do not collude. 

\textbf{STIP insights and design.}
Based on our introduced three-party threat model, we developed STIP, Secure Transformer Inference Protocol, with two insights.
First, we employ efficient feature-space permutation for secure and equivalent Transformer inference.
Since the inference is executed on the untrusted server, model parameters and on-device data must be transformed before uploading to the cloud.
Based on efficient permutation in the feature space, we design a data and parameter transformation approach for Transformer layers.
We prove the mathematical equivalence of computation using our proposed transformation, thus ensuring no loss of accuracy.
Second, we design a semi-symmetrical protection scheme between the model developer and data owners.
This insight stems from the sequential structure of neural networks.
We reveal that the model developer only needs to share identical permutations in the first and last layers with the data owner, and can exclusively retain the information of intermediate layer transformation.
Similar semi-symmetric protection schemes have found application in diverse domains, such as image encryption~\cite{di2017semi} and online shopping~\cite{fares2016novel}.
We demonstrate the privacy-preserving capability of STIP based on distance correlation~\cite{distcorr} and prove its resistance to brute-force and known-plaintext attacks.

\textbf{Contributions.}
We summarize three key contributions of this work as follows:

\noindent \textbullet~We identify the efficiency bottleneck inherent in the two-party setting and its misalignment with real-world applications. 
We propose a new three-party threat model, decomposing the model owner into two distinct entities: model developer and model server. 

\noindent \textbullet~We present STIP, the first secure protocol for three-party Transformer inference, with the theoretical bound of privacy leakage and guarantee of no loss of accuracy.

\noindent \textbullet~We implement STIP and conduct evaluations on various Transformer models with up to 70 billion parameters on real systems. 
Experimental results showcase the efficiency of STIP, comparable to unprotected full-cloud inference, surpassing state-of-the-art secure two-party protocols~\cite{hou2023ciphergpt, hao2022iron, chen2022x} by millions of times.

\noindent \textbf{This work does not raise any ethical issues.}
\vspace{-0.15in}
\section{Background}

This section first introduces device-cloud collaboration ($\S$~\ref{subsec:dc-collab}) and prior efforts for secure Transformer inference with the two-party setting ($\S$~\ref{subsec:secure-twop}). 
From experience of managing real-world Transformer-based services, we propose a three-party threat model ($\S$~\ref{subsec:threep}) and claim the scope, design goals, and uniqueness of our secure protocol ($\S$~\ref{subsec:dspace}).

\subsection{Device-Cloud Collaboration}
\label{subsec:dc-collab}

Transformer inference-based services have become the most compelling artificial intelligence applications, e.g., ChatGPT sets a record for the fastest-growing user base~\cite{chatgpt-record}.

\textbf{Full-cloud inference: efficient but unsafe.}
Serving Transformers, especially large models with billions of parameters, aligns with cloud computing capabilities, like auto-scaling~\cite{romero2021infaas}.
Several cloud-native frameworks for Transformer inference have been released, like NVIDIA NeMo~\cite{kuchaiev2019nemo} and Microsoft DeepSpeed~\cite{aminabadi2022deepspeed}.
Therefore, most organizations, including OpenAI, adopt full-cloud deployment for their Transformer-based services~\cite{azure-openai}.
However, sending raw data to the cloud is infeasible for various privacy-sensitive areas, e.g., Samsung officially banned employees from using ChatGPT after sensitive code was leaked~\cite{samsung}.

\textbf{Full-device inference: secure but not scalable.}
To address data privacy concerns, another option is deploying Transformer models entirely on the device.
Through model compression techniques like weight quantization~\cite{llmint8}, Transformer models with billions of parameters can run inference on devices~\cite{gptondevice}.
However, the scalability of full-device deployment is limited.
First, FLOPs (floating point operations) and memory footprints scale linearly with the number of parameters, while on-device computing resources grow much slower than the explosive size of Transformer models~\cite{llmsurvey, ondeviceDL}.
Second, model compression inevitably brings accuracy loss~\cite{xiao2023smoothquant}, but in the highly competitive large Transformer market, even a slight QoS degradation may result in lagging behind competing products.~\cite{Open-LLM-Leaderboard-Report-2023}.

\textbf{Device-cloud collaboration: best of both worlds.}
Since neither the cloud nor the device can satisfactorily serve Transformer models independently, device-cloud collaboration has naturally emerged in recent research efforts~\cite{dc-transformer, thapa2022splitfed, zeng2020coedge}.
By reasonably assigning inference workloads between the device and the cloud, collaborative inference avoids transmitting raw data while maintaining efficiency~\cite{pham2023binarizing, pasquini2021unleashing}.

\subsection{Secure Two-Party Inference}
\label{subsec:secure-twop}

Not transmitting raw data is just a bottom line for security.
Research on attacks against deep neural networks shows that intermediate activations (e.g., text embeddings~\cite{pan2020privacy}) can be reverse-engineered to disclose the sensitive information in raw data~\cite{abuadbba2020can}.
Tighter security is urgently needed.

\textbf{HE and 2PC for Transformer inference.}
Homomorphic encryption (HE) is a cryptographic technique that allows computations to be performed on encrypted data without decrypting it, while multi-party computation (MPC) allows multiple parties to jointly compute a function over their inputs while keeping those inputs private. 
In the context of model inference, two-party computation (2PC) is usually considered, a special case of MPC where the model owner (cloud server) and the data owner (device) represent two parties respectively.
Recent research has demonstrated the ability to serve Transformer inference using a combination of HE and 2PC~\cite{hou2023ciphergpt, hao2022iron, chen2022x}.
However, these protocols incur significant computational overhead when processing non-linear complex layers, such as LayerNorm and ReLU. 
Additionally, there are high costs associated with device-cloud communications, e.g., CipherGPT takes over 25 minutes to generate a single token with GPT2 model~\cite{hou2023ciphergpt}.

\subsection{Align with Real-World Applications: Three-Party Threat Model}
\label{subsec:threep}

The simplicity of the two-party setting, where one party represents the device and the other the cloud, seamlessly fits HE and 2PC theories.
However, the efficiency challenges also stem from the computational hardness inherent in HE and MPC theories~\cite{mpc-complexity}, which motivates us to think about a question:
\textit{Does the two-party setting truly align with the demands of real-world applications?}

Surprisingly but fortunately, the answer is no.
This conclusion comes from our experience of managing two real-world Transformer-based services.

\textbf{Service-1: campus security chatbot.}
At ANONYMOUS University, we host a large language model-based chabot for campus security.
Our chatbot uses the database of surveillance video analytics as the information source.
Users, including students and campus security officers, can ask questions like ``Did any abnormal behavior occur during a certain period?'' to the chatbot and get responses in natural language.

\textbf{Service-2: vehicle cabin assistant.}
At ANONYMOUS automotive company, we deploy a multi-modal Transformer model to enhance the functionality of the smart assistant in vehicle cabins.
The multi-modal Transformer takes the in-cabin video frames as the input and generates scene descriptions in natural language.
Scene descriptions can help the in-car assistant become more user-friendly, such as recommending music based on facial expressions.

\textbf{Common experience: the model developer is not the model server.}
For both services, we developed the Transformer model by fine-tuning~\cite{lora} open-sourced parameters~\cite{touvron2023llama, vit} with our collected data.
The computing power of our on-campus lab and company can afford offline model development, but cannot support large-scale long-term services.
There are approximately a few hundred users of the campus chatbot, and the vehicle assistants need to serve hundreds of thousands of users.
We, as model developers, need to resort to third-party cloud computing platforms to serve our Transformer models.
In fact, this experience is not just for us, but also for other model development companies such as OpenAI. 
OpenAI uses the Microsoft Azure cloud platform to provide ChatGPT services to hundreds of millions of users~\cite{chatgpt-cloud, azure-openai}.

\begin{figure}[t]
    \centering
    \begin{subfigure}[b]{1\linewidth}
        \centering
        \includegraphics[width=0.9\linewidth]{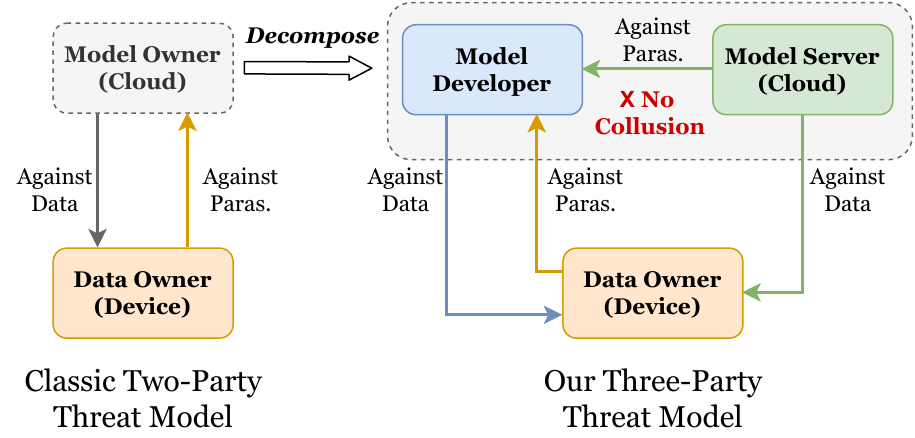}
    \end{subfigure}
    \caption{Three-Party Threat Model}
    \vspace{-0.2in}
    \label{fig:threeparty}
\end{figure}

\textbf{Three-party threat model.}
To align with experience from developing real-world services, we introduce a three-party threat model, as shown in Fig.~\ref{fig:threeparty}.
Departing from the classic two-party setting, we divide the model owner into two distinct entities: the model developer and the model server.
Given that the developed models are proprietary, model developers must safeguard their model parameters against attacks from model servers~\cite{model-confidential}.
We make the assumption that model developers do not engage in collusion with cloud platforms. 
Any collaboration between these entities would lead to a regression to the classic two-party setting.
This decomposition of the model owner role not only enhances practicality but also relaxes adversary assumptions.
This adjustment proves instrumental in overcoming the computational hardness inherent in two-party protocols.

It is worth noting that our three-party setting also has limitations. 
For example, it is not perfectly suitable for cloud providers that can also develop their own models, such as Google Gemini~\cite{gemini}. 
For more discussion, please see Sec.~\ref{sec:discuss}.

\subsection{Design Space}
\label{subsec:dspace}

\textbf{Scope.}
This work focuses on achieving efficient and secure execution of Transformer model inference, i.e., the end-to-end forward pass, within the three-party setting. 
Transformer inference results can serve as a foundation for various downstream applications, such as AI agents that can automatically call external APIs~\cite{ruan2023tptu}.
Potential security risks and efficiency issues in downstream applications are out of the scope of this work.

\textbf{Design goals.}
Our protocol has four main design goals:

\noindent \textit{(1) Data and parameter security.}
The foremost objective is to ensure the security of on-device data and model parameters.

\noindent \textit{(2) No accuracy loss.}
The protocol is required to perform accuracy-lossless inference, meaning there should be no approximation of any computations in Transformer models.

\noindent \textit{(3) Support production environments.}
It is crucial that the protocol supports inference frameworks used in production environments, incorporating techniques such as kv-cache for efficiency optimization~\cite{aminabadi2022deepspeed, kuchaiev2019nemo}.

\noindent \textit{(4) Flexible extension to Transformer variants.}
Given the continuous evolution of Transformer models with various emerging variants~\cite{touvron2023llama, gpt2, liu2023llava, jiang2024mixtral, longformer}, the protocol must possess the ability for flexible extension to accommodate Transformer variants. This ensures long-term availability without necessitating case-by-case adaptation.

\begin{table}[t]
\centering
\caption{Comparison of our proposed STIP and existing Transformer inference methods.}
\label{tab:comparison}
\begin{tabular}{@{}lcccc@{}}
\toprule
Method & Scr. & \begin{tabular}[c]{@{}c@{}}No\\ Loss\end{tabular} & \begin{tabular}[c]{@{}c@{}}Prd.\\ Env.\end{tabular} & \begin{tabular}[c]{@{}c@{}}Tested\\ Models\end{tabular} \\ \midrule
Full-Cloud & \xmark & \cmark & \cmark & All \\ \midrule
Iron~\cite{hao2022iron} & \cmark & \xmark & \xmark & BERT series \\
THE-X~\cite{chen2022x} & \cmark & \xmark & \xmark & BERT-Tiny \\
CipherGPT~\cite{hou2023ciphergpt} & \cmark & \xmark & \xmark & GPT-2 \\ \midrule
STIP & \cmark & \cmark & \cmark & \begin{tabular}[c]{@{}c@{}}GPT/LLaMA/\\ ViT/LLaVA/\\ BERT/Mixtral series\end{tabular} \\ \bottomrule
\end{tabular}
\end{table}

Tab.~\ref{tab:comparison} shows the comparison of \mysys and existing Transformer inference methods on our four design goals.

\section{Challenges}
In designing STIP that achieves both efficiency and security, we encountered two non-trivial challenges.

\subsection{Prohibitive Cryptographic Overhead}
\label{subsec:c1-overhead}

\begin{figure}[t]
    \centering
    \begin{subfigure}[b]{\linewidth}
        \centering
        \includegraphics[width=0.9\linewidth]{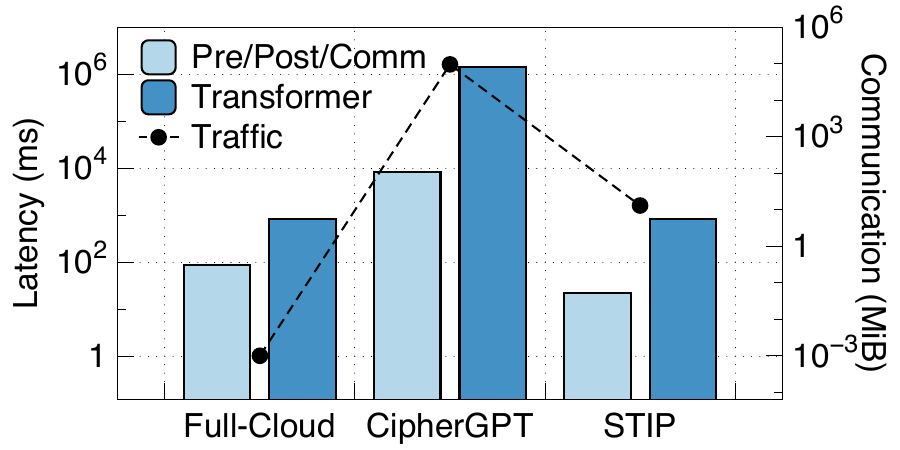}
    \end{subfigure}
    \caption{Latency and communication overheads of GPT2-124m model.
    }
    \vspace{-0.1in}
    \label{fig:challenge-overhead}
\end{figure}

Transformation of data and parameters is key to protection.
Existing protocols that combine HE and 2PC techniques for security have prohibitive computing and communication overheads.
As shown in Fig.~\ref{fig:challenge-overhead}, CipherGPT~\cite{hou2023ciphergpt} takes over 25 minutes and 90 GiB traffic to perform a single forward pass of GPT2~\cite{gpt2} with 123 million parameters.

The success of the Transformer model hinges on the utilization of global matrix multiplication in its self-attention and feedforward modules, making it highly parallelizable compared to recursive architectures~\cite{rush2018annotated}. 
The inductive bias of the Transformer architecture is not only efficient at the implementation level but also has some properties such as permutation symmetries of hidden units~\cite{ainsworth2022git} and the token-wise permutation invariance~\cite{lee2019set}. 
Drawing attention to the prevalent use of random permutation in addressing privacy and security concerns, including secure communication~\cite{bhargava2017new}, property inference~\cite{ganju2018property}, machine learning~\cite{maekawa2018privacy, secure-ML-permute, he2020transnet}, etc.
In light of these observations, we present our first insight:
\myhighlight{Insight 1: Efficient feature-space permutation for transformed and equivalent Transformer inference.}
Since the inference is executed on the untrusted server, the parameters from the model developer and the on-device data must be transformed before uploading to the cloud.
We design a data and parameter transformation specifically for Transformer layers based on random permutation in the feature space.
The transformation can be efficiently implemented with $O(d)$-complexity movement of memory pointers, where $d$ is the feature dimension.
As Fig.~\ref{fig:challenge-overhead} shows, our protocol, \mysys, achieves orders of magnitude higher efficiency than CipherGPT, and the latency is close to full-cloud deployment.
We prove the mathematical equivalence of computation using our proposed transformation, thus ensuring no loss of accuracy ($\S$~\ref{subsec:feat-space}).

\subsection{Attack Surface Vulnerability}
\label{subsec:c2-security}

While random permutation-based schemes are efficient and accuracy-lossless, endowing them with robust security poses non-trivial challenges.
A direct adoption of sequence-level permutation scheme~\cite{lee2019set} leads to $n!$ possible permutations, rendering it inadequate for safeguarding data against brute-force attacks (BFA) when the number of input tokens $n$ is small.
Opting for permutation in the feature space can enhance protection, yet using a single permutation matrix $\pi$ remains vulnerable to known-plaintext attacks (KPA).
The reason is, once the cloud gains a pair of known plaintext of on-device data and the transformed data, it can easily recover the permutation matrix, subsequently inverse-transforming all parameters and exposing sensitive information.
Tab.~\ref{tab:challenge-secure} summarizes the number of possible permutations for data and parameters and resistance to BFA and KPA attacks.
\myhighlight{Insight 2: Semi-symmetrical set of permutation matrices between the model developer and data owners.}
This insight stems from the sequential structure of neural networks.
The data owner only needs to share identical permutations in the first and last layers with the model developer. 
Intermediate layer transformation information can be exclusively retained by the model developer.
Consequently, we propose a feature-space permutation scheme utilizing a set of matrices ${\pi_1, ..., \pi_{3L}}$, where $L$ represents the number of layers. 
Similar semi-symmetric protection schemes have been explored in domains such as image encryption~\cite{di2017semi} and online shopping~\cite{fares2016novel}.
We analyze the privacy-preserving capability of our proposed scheme based on distance correlation~\cite{distcorr} and prove its resistance to BFA and KPA ($\S$~\ref{subsec:sec-ana}).

\begin{table}[t]
\centering
\caption{Comparison of different permutation-based schemes in terms of number of possibilities and resistance to attacks.}
\label{tab:challenge-secure}
\begin{tabular}{@{}l|cc|cc@{}}
\toprule
Protection Scheme & Data & Paras. & BFA & KPA \\ \midrule
Seq. Perm. & $n!$ & 1 & \xmark & \xmark \\
Feat. Perm. with Single $\pi$ & $d!$ & $d!$ & \cmark & \xmark \\
Feat. Perm. with $\{\pi_1, ..., \pi_{3L}\}$ & $d!$ & $(d!)^{3L}$ & \cmark & \cmark \\ \bottomrule
\end{tabular}
\end{table}

\section{Definition}
\label{sec:problem}

With these insights in mind, in order to formally present mathematically equivalent transformations and analyze theoretical security, this section defines Transformer inference ($\S$~\ref{subsec:trn-inf}) and the three-party threat model ($\S$~\ref{subsec:three-party}).

\subsection{Transformer Inference}
\label{subsec:trn-inf}

We use the original Transformer architecture~\cite{vaswani2017attention} to introduce the inference workflow, without loss of generality, advanced Transformer variants (GPT\cite{radford2019language}, LLaMA~\cite{touvron2023llama}, ViT~\cite{dosovitskiy2020image} and Mixtral~\cite{jiang2024mixtral}) will be discussed in Sec.~\ref{sec:real}.

\begin{figure}[t]
    \centering
    \begin{subfigure}[b]{0.8\linewidth}
        \centering
        \includegraphics[width=\linewidth]{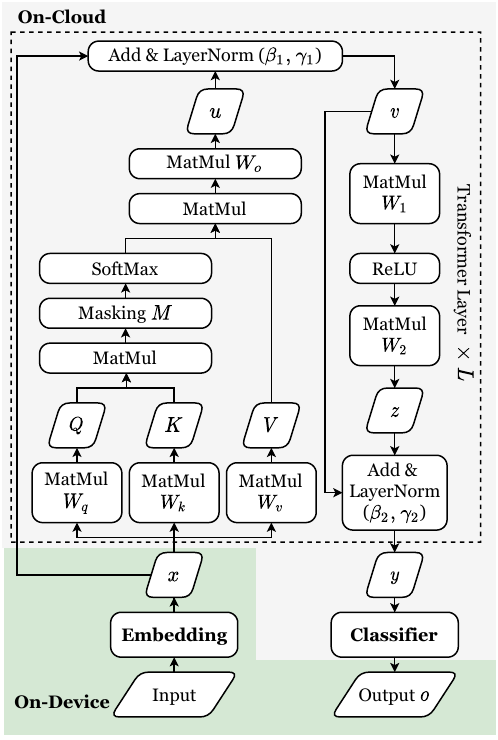}
    \end{subfigure}
    \caption{Inference workflow of original Transformer. 
    }
    \vspace{-0.1in}
    \label{fig:trn-arch}
\end{figure}

\textbf{Device-cloud model split}
In Transformer models, the embedding operation is the initial step that maps discrete inputs (such as words or images) into continuous vectors~\cite{vaswani2017attention}.
In order to decouple our protocol from the original data modality, we set the starting point of device-cloud collaborative inference to the embeddings instead of the input.
By default, only the embedding operation executes on the device while expensive Transformer layers and the classifier are deployed on the cloud, as shown in Fig.~\ref{fig:trn-arch}.


\textbf{Transformer layer forward pass.}
As shown in Fig.~\ref{fig:trn-arch}, the on-cloud Transformer model consists of $L$ sequential Transformer layers and a classifier. 
Let $F_\theta$ denote the Transformer model with trainable parameters $\theta$.
We define $\{f_i:\mathbb{R}^{n\times d} \mapsto \mathbb{R}^{n\times d}| i\in [L]\}$ as Transformer layers, and $f_c:\mathbb{R}^{n\times d} \mapsto \mathbb{R}^{n\times s}$ as the classifier, where $n$ is the sequence length (e.g., the number of tokens), $d$ is the model feature dimension, and $s$ is the output dimension (e.g., vocabulary size).
We use $x_i$ and $y_i$ to denote the input and output of the $i$-th Transformer layer, and all these intermediate activations share the same shape $\mathbb{R}^{n \times d}$.
A forward pass of a Transformer layer, i.e., $f(x)=y$, is computed as follows\footnote{For simplicity of expression, we use $xW$ instead of $xW^T$ which is used for real-world implementation.}: 

\noindent Self-attention sub-block:
\begin{align*}
    &Q = xW_q,~~K = xW_k,~~V = xW_v, &W_q, W_k, W_v \in \mathbb{R}^{d\times d},\\
    &u = \text{SoftMax}\left( \frac{QK^T}{\sqrt{k}} + M\right)V W_o, & M \in \mathbb{R}^{n\times n}, W_o \in  \mathbb{R}^{d\times d},\\
    &v = \text{LayerNorm}(u + x; \gamma_1, \beta_1), &\gamma_1, \beta_1 \in \mathbb{R}^d,
\end{align*}

\noindent Feedforward sub-block:
\begin{align*}
    &z = \text{ReLU}(v W_1)W_2, &W_1\in \mathbb{R}^{d\times m},W_2\in \mathbb{R}^{m\times d},\\
    &y = \text{LayerNorm}(z + v; \gamma_2, \beta_2), &\gamma_2, \beta_2 \in \mathbb{R}^d,
\end{align*}
where $k$ and $m$ are constants that depend on the model architecture hyperparameters, and $M$ denotes the mask matrix.
SoftMax, LayerNorm, and ReLU are commonly used neural network functions and their definitions are not necessary in this work.
Following the $L$-layer Transformers, a classifier computes as follows:
\begin{align*}
    &o = \text{SoftMax}\left(y_L W_c\right),  & W_c \in \mathbb{R}^{d\times s}.
\end{align*}

\textbf{Use of masking.}
The mask matrix is a lower triangular matrix, where the elements above the main diagonal are set to negative infinity, and the elements on and below the main diagonal are set to zero.
In the original Transformer work~\cite{vaswani2017attention}, two types of Transformer layers are proposed, encoder and decoder.
The mask is only applied to the self-attention sub-block in the decoder to prevent positions after the current position from being attended to. 
It ensures that during the generation of each token in the output sequence, the model only attends to the tokens preceding it in the sequence. 
Masking operation is not trivial, as it results in the infeasibility of constructing equivalent computations using sequence-level permutation ($\S$~\ref{subsec:feat-space}).


\subsection{Three-Party Setting and Threat Model}
\label{subsec:three-party}

For serving Transformer models, we consider three parties:

\noindent \textbullet~Model developer ($P_1$) that trains and owns private Transformer model parameters, e.g., OpenAI developed GPT4. 

\noindent \textbullet~Model server ($P_2$) that has the computing hardware, e.g., high-end GPUs on Azure cloud platform.

\noindent \textbullet~Data owner ($P_3$) that own private input and inference output, e.g., text prompts and responses of ChatGPT users. 

\noindent The inference protocol should ensure that $P_1$ and $P_2$ are unaware of $P_3$'s input $x_1$ and inference output $o$.
And $P_1$'s model parameters $\theta$ should be hidden to both $P_2$ and $P_3$.
The parameter $\theta$ consists of attention weights ($W_q, W_k, W_v, W_o$), feedforward weights ($W_1, W_2$), LayerNorm weights ($\gamma, \beta$), and classifier weights ($W_c$).
In our context of Transformer inference, $P_3$'s input can be text prompt~\cite{touvron2023llama}, images~\cite{vit}, and a combination of multiple modalities~\cite{liu2023llava}, and the inference output is typically a probability vector of the last classification head~\cite{llmsurvey}.

We adopt the widely used semi-honest setting~\cite{hou2023ciphergpt, hao2022iron, secure-ML-permute, chen2022x} where each party will follow the protocol specifications but attempt to infer additional sensitive information from the observed protocol messages.

\section{Design}
\label{sec:design}

This section first introduces how to perform equivalent inference of Transformer models with feature space permutation ($\S$~\ref{subsec:feat-space}).
Then we present, Secure Transformer Inference Protocol (\mysys), the core algorithm ($\S$~\ref{subsec:protocol}), and analyze the protocol security ($\S$~\ref{subsec:sec-ana}).

\subsection{Feature Space Permutation}
\label{subsec:feat-space}

The permutation operation is defined by a permutation matrix $\pi$, which is a square binary matrix that has exactly one entry of 1 in each row and each column with all other entries of 0.
For $x \in \mathbb{R}^{n \times d}$, $\pi_{n\times n} x$ and $x \pi_{d\times d}$ performs sequence-level and feature-level permutation respectively.

\textbf{Mask makes sequence-level permutation not equivalent.}
For the Transformer encoder (self-attention without masking), the sequence-level permutation equivariance property, i.e., $f(\pi x)=\pi f(x)$, has been studied~\cite{lee2019set}.
However, due to the mask inside the decoder, attention computation on sequence-level permuted data cannot return equivalent output.
A quick fix is to send a permuted $M'=\pi M \pi^T$ to the cloud computing platform.
However, since the value structure of $M$ is known, the cloud platform can easily infer $\pi$, which will result in a loss of protection.

\begin{figure}[t]
    \centering
    \begin{subfigure}[b]{\linewidth}
        \centering
        \includegraphics[width=\linewidth]{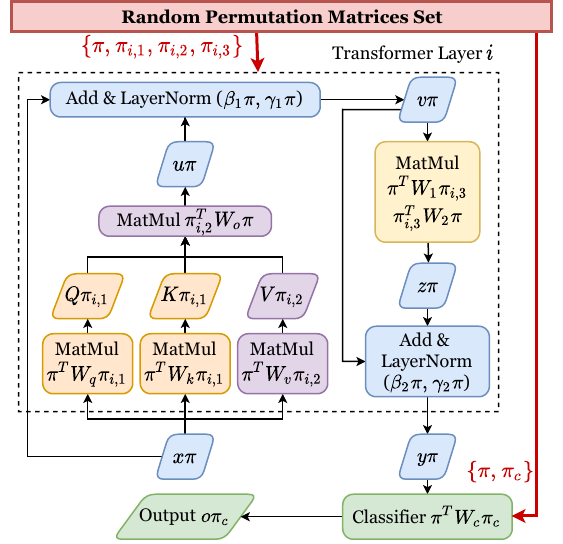}
    \end{subfigure}
    \caption{Illustration of feature-space parameter transformation. Colors represent the use of different permutation matrices.}
    \label{fig:paratrans}
\end{figure}

\textbf{Parameter transformation.}
Instead, we propose to transform parameters in the feature space with a set of random permutation matrices.
First, we generate $\pi \in \{0,1\}^{d\times d}$ for the input $x$.
For the $i$-th Transformer layer, we transform parameters with another three matrices $\pi_{i,1}, \pi_{i,2}, \pi_{i,3}$:
\begin{align*}
    &W_q' = \pi^T W_q \pi_{i,1},~~W_k' = \pi^T W_k \pi_{i,1},~~W_v' = \pi^T W_v \pi_{i,2},\\
    &W_o' = \pi_{i,2}^T W_o \pi,~~W_1' = \pi^T W_1 \pi_{i,3},~~W_2' = \pi_{i,3}^T W_2\pi,\\ &\gamma_1'=\gamma_1\pi,~~\beta_1'=\beta_1\pi,~~\gamma_2'=\gamma_2\pi,~~\beta_2'=\beta_2\pi.
\end{align*}
For the classifier, we need to generate a permutation matrix $\pi_c \in \{0,1\}^{s\times s}$.
We transform the classifier parameters by
\begin{align*}
    &W_c' = \pi^T W_c \pi_c.
\end{align*}
Fig.~\ref{fig:paratrans} illustrates the parameter transformation procedure.

\textbf{Computational equivalence.}
Let $F_{\theta'}$ denote the Transformer model with transformed parameters, we prove that original inference results can be recovered equivalently:
\begin{theorem}
\label{thm:1}
    $F_{\theta '}(x\pi)\pi_c^T = F_{\theta}(x)$.
\end{theorem}
\noindent Due to page limitations, the proof is placed in Appendix~\ref{appendix:proof}.

\subsection{Protocol}
\label{subsec:protocol}

\begin{figure}[t]
    \centering
    \begin{subfigure}[b]{0.9\linewidth}
        \centering
        \includegraphics[width=\linewidth]{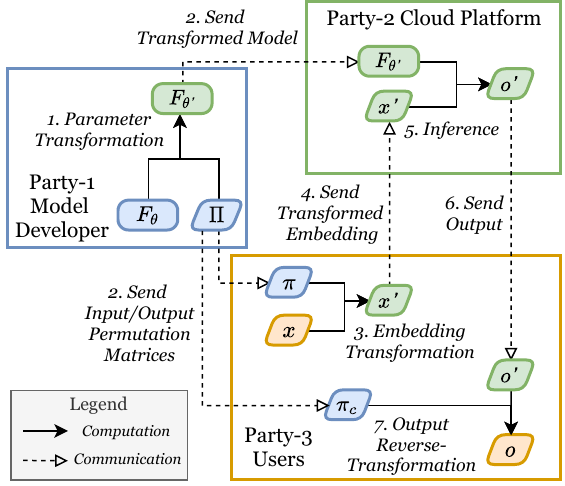}
    \end{subfigure}
    \caption{Overview of STIP.}
    \vspace{-0.2in}
    \label{fig:overview}
\end{figure}

Based on our proposed permutation-based transformation for Transformer models, we develop \mysys protocol.
Fig.~\ref{fig:overview} shows an overview of STIP.
\mysys has two phases: initialization and inference.
In the initialization phase, the model developer $P_1$ randomly generates the permutation matrices set $\Pi=\{\pi, \pi_c \}\cup\{\pi_{i,1},\pi_{i,2},\pi_{i,3} | i\in [L]\}$.
$P_1$ transforms its owned trained model $F_\theta$ with $\Pi$ and obtain the transformed version $F_{\theta'}$.
Then $P_1$ sends the transformed model $F_{\theta'}$ to the cloud platform and distributes the permutation matrices for the input and output ($\pi$ and $\pi_c$) to its registered users.
Now the initialization phase finishes.
For the inference phase, once a user wants to use the inference service, it runs the embedding on the device to obtain $x$.
Then the user transforms the embedding using the received input permutation matrix $\pi$ by a super-lightweight operation $x\pi=x'$.
Then the user sends $x'$ to the cloud.
The workload on the cloud platform has no change, compared with the normal Transformer model serving.
The cloud just performs the $F_{\theta'}(x')$ computations and obtains the output $o'$ in the permuted feature space.
Once the user receives the returned $o'$ from the cloud, it simply reverse-transform the output by $o=o'\pi_c^T$, which involves only memory movement operations and can be implemented super efficiently.
Till now, one round of inference finished.
Alg.~\ref{alg:stip} formally presents \mysys protocol.

\textbf{Multi-pass autoregressive generation.}
The service of modern language models is based on autoregressive generation, wherein the model predicts tokens sequentially, one token at a time, with each token being conditioned on the given prompt and previously generated ones.
\mysys can support autoregressive generation services by requiring the user to combine locally predicted tokens with its initial prompt and subsequently re-embedding the input.
Unlike full-cloud inference which involves a single round of communication, STIP protocol introduces $n$ rounds of communication, where $n$ is the number of tokens finally generated.
However, this additional communication overhead is inevitable due to the requirement of preserving the confidentiality of inference output. 
Our experiments show that \mysys can generate 100 tokens in the autoregressive mode in 3s latency with a wired connection.
Considering that full-cloud inference has over 1s delay, \mysys is practically efficient while providing critical security guarantees.

\textbf{Production environment support.}
Production-level Transformer inference frameworks, such as DeepSpeed~\cite{aminabadi2022deepspeed} and HuggingFace~\cite{wolf2020huggingfaces}, incorporate many techniques to enhance efficiency.
For example, they use KV-cache mechanism to alleviate redundant computations by storing intermediate results from previous attention calculations.
\mysys maintains a non-intrusive approach, transforming parameter values while keeping the underlying Transformer architecture unchanged. 
From the cloud perspective, \mysys involves switching to a distinct set of weights, with no change in the inference computation process. 
As a result, \mysys seamlessly aligns with production-level frameworks, and we have implemented \mysys with the HuggingFace~\cite{wolf2020huggingfaces} library.

\begin{algorithm}[t]
\caption{Secure Transformer Inference Protocol}
\label{alg:stip}
\SetKwInOut{Input}{input}
\SetKwFunction{Send}{Send}
\SetKwFunction{ParaTrans}{ParaTrans}
\Input{Number of Transformer Layers $L$}

\textbf{Initialization phase:}\\
~~[$P_1$] $\Pi \gets \{\pi, \pi_c \}\cup\{\pi_{i,1},\pi_{i,2},\pi_{i,3} | i\in [L]\}$\;
~~[$P_1$] $F_{\theta'} \gets$ \ParaTrans($F_{\theta}, \Pi$)\;
~~[$P_1$] \Send($F_{\theta'}$, $P_2$) and \Send($\{\pi,\pi_c\}$, $P_3$)\;
\textbf{Inference phase:}\\
~~[$P_3$] $x'\gets x\pi$ and \Send($x'$,$P_2$)\;
~~[$P_2$] $o'\gets F_{\theta'}(x')$ and \Send($o'$, $P_3$)\;
~~[$P_3$] $o \gets o'\pi_c^T$.
\end{algorithm}

\subsection{Security Analysis}
\label{subsec:sec-ana}

Now we demonstrate that \mysys can protect model parameters and user data from various attacks and quantify the bound of privacy leakage risk using distance correlation measure. 

\textbf{Random permutation resists brute-force attacks.}
To begin, let's consider $P_1$ as the attacker attempting to access user data $x,o$. 
Due to the inaccessible nature of $x\pi$ and $o\pi_c$, $P_1$ is unable to recover $x, o$ with possessing $\pi, \pi_c$.
Next, consider $P_2$ as the attacker targeting model parameters $\theta$ and user data $x,o$. 
Given that $P_2$ possesses permuted parameters and $x\pi$, the probability of correctly guessing $\pi_{d\times d}$ is $1/(d!)$, where $d$ is typically larger than 512 in practical applications such as $d=4096$ in LLaMA2-7b~\cite{touvron2023llama}. 
This renders the likelihood of a successful attack negligible. 
Notably, permutation-based protection schemes often exhibit a weakness in preserving the set of elements (e.g., English vocabulary)~\cite{secure-ML-permute}. 
Fortunately, \mysys avoids this vulnerability by applying permutation to intermediate activations rather than the original data.
Thirdly, consider $P_3$ as the attacker against model parameters $\theta$. 
Since $P_3$ lacks access to $\theta'$, it cannot recover $\theta$ despite having $\pi, \pi_c$. 
As \mysys requires deploying the embedding model on the device, the weights of the embedding are exposed to $P_3$. 
However, the embedding module alone cannot perform valuable tasks and is therefore not sensitive (e.g., OpenAI has released its embedding module~\cite{tiktoken}).

\textbf{Semi-symmetrical scheme to resist known-plaintext attack.}
A known-plaintext attack (KPA) is a cryptographic attack where the adversary possesses both the ciphertext (encrypted data) and the corresponding plaintext (unencrypted data). 
The goal of KPA is to uncover the encryption key or algorithm used to encrypt the data. 
In our context, if the plaintext of the model developer's parameters has been leaked, there is no need to continue attacking the protection scheme.
Therefore, the focus of KPA consideration lies exclusively on user data.
Assuming $P_2$ knows both $x$ and $x\pi$, it can recover $\pi$ with $d$ times column matching, unless there are two or more exactly identical columns.
Consequently, if parameters on the cloud rely solely on $\pi$ for protection, all of them are at risk of being leaked.
This underscores the rationale behind our adoption of a semi-symmetric protection scheme, wherein layer parameters are permuted using two matrices.
One is exclusively owned by the model developer, while the other is shared with the user. 
This design in \mysys makes the model parameters resistant to KPA.
For a specific user, uncovering $\pi$ would lead to all subsequent embeddings being exposed to $P_2$.
To address this vulnerability, we implement a strategy of periodically changing the set of permutation matrices (in extreme cases, using one-time transformation), a practice commonly employed to resist KPA~\cite{zhao2016information, bianchi2015analysis}.

\textbf{Social engineering attack.}
Setting aside our semi-honest assumption, in a scenario where the cloud platform deceitfully pretends as a user and acquires the embedding model along with $\pi, \pi_c$, it can potentially uncover the data of other users who share the same permutation matrices.
To counteract this risk, the model developer can deploy multiple instances of the model, each employing distinct transformations. 
Users can then be randomly assigned to share a model instance (in extreme cases, each user may have an exclusive model instance), effectively mitigating the risk of data leakage through this social engineering attack.
It's noteworthy that parameters remain resistant to this attack for the same reasons as the KPA we discussed above.

\textbf{Distance correlation bound.}
Above we analyzed that \mysys can ensure that data values cannot be uncovered.
An important aspect to investigate is the degree of correlation between the original and permuted data. 
To quantify the risk of privacy leakage, we employ distance correlation~\cite{distcorr}, a statistical measure of dependency between two random vectors.
Let Corr denote the distance correlation.
Based on existing theorem~\cite{secure-ML-permute}, it has been proven that:
\begin{equation*}
    \mathop{\mathbb{E}}\limits_{\substack{\pi_{d\times d}, A\in \mathbb{R}^{d\times d}}} [\text{Corr}(x, xA\pi)] \leq \mathop{\mathbb{E}}\limits_{B\in \mathbb{R}^{d\times 1}} [\text{Corr}(x, xB)].
\end{equation*}
In simpler terms, this implies that the privacy leakage resulting from the application of random permutation on intermediate activations is bounded by the leakage observed in one-dimensional reduction, which has demonstrated practical privacy-preserving capabilities~\cite{oliveira2004privacy, wang2018theoretical}.

\textbf{Model split considerations.}
By default, \mysys only deploys the embedding module on user devices, as shown in Fig.~\ref{fig:trn-arch}.
This decision is motivated by the fact that splitting the model before the embedding poses security challenges. 
In such a scenario, the device would be required to transmit tokenized one-hot vectors (a matrix $\in \mathbb{Z}^{n\times s}$, where $s$ denotes the vocabulary size) to the cloud. 
While the matrix can be randomly permuted, the inherent one-hot nature of the vectors makes it susceptible to easy recovery of the permutation by the cloud, rendering it insecure.
On the flip side, distributing more layers onto the device is also not a prudent choice. 
This is primarily because exposing additional parameters to end devices not only compromises efficiency, as evidenced by experimental results in Fig.~\ref{fig:splitlat}, but also fails to enhance the model's theoretical resistance to cloud-device attacks. 
\section{STIP for Transformer Variants}
\label{sec:real}

This section discusses how \mysys supports various Transformer variants, including language models ($\S$~\ref{subsec:real-lan}), multi-modal models ($\S$~\ref{subsec:real-mm}), and mixture-of-experts models ($\S$~\ref{subsec:real-moe}).
Following that, we establish generalized rules and claim the application scope of \mysys ($\S$\ref{subsec:real-scope}).

\noindent \textit{Due to page limitations, all proofs are placed in Appendix.~\ref{appendix:proof}.}

\subsection{Language Models}
\label{subsec:real-lan}

\textbf{Pre-LayerNorm.}
The first version of GPT~\cite{gpt} directly adopts the original Transformer decoder.
GPT-2~\cite{gpt2} introduces Pre-LayerNorm, relocating layer normalization to the input of self-attention and feedforward sub-blocks, formally:
\begin{align*}
    v &= \text{Attn}(\text{LayerNorm}(x))+x,\\
    y &= \text{ReLU}(\text{LayerNorm}(v) W_1)W_2 + v,
\end{align*}
where Attn denotes the self-attention sub-block.
From the proof of Theorem~\ref{thm:1}, we can easily prove that this theorem still holds for the Pre-LayerNorm structure, using the same parameter transformation.

\textbf{RMSNorm.}
LLaMA series~\cite{touvron2023llama} use RMSNorm~\cite{zhang2019root} to replace LayerNorm.
To support RMSNorm operator, \mysys transforms its weight $\gamma$ by $\gamma \pi$, then we can prove that $$\text{RMSNorm}(x\pi;\gamma\pi)=\text{RMSNorm}(x;\gamma)\pi.$$

\textbf{GeLU.}
GPT uses GeLU to replace ReLU in feedforward sub-blocks.
Analogous to ReLU, GeLU involves element-wise computation without learnable parameters, hence we have $\text{GeLU}(x\pi)=\text{GeLU}(x)\pi$ and theorem~\ref{thm:1} holds.

\textbf{SwiGLU feedforward.}
LLaMA~\cite{touvron2023llama} uses SwiGLU~\cite{shazeer2020glu} instead of ReLU in feedforward layers.
Let FFN$_{\text{SwiGLU}}$ denote the feedforward layers using SwiGLU, with the definition:
\begin{align*}
    \text{FFN}_{\text{SwiGLU}}(x) &= \left( xW_1\text{Sigmoid}(xW_1)xW_3\right)W_2,\\ 
    &W_1, W_3 \in \mathbb{R}^{d\times m}, W_2 \in \mathbb{R}^{m\times d}.
\end{align*}
We transform parameters as follows:
\begin{align*}
    W_1' = \pi^T W_1,~~W_3' = \pi^T W_3 \pi_{i, 3},~~W_2'=\pi_{i,3}^T W_2\pi,
\end{align*}
where FFN$'_{\text{SwiGLU}}$ denote the transformed feedforward sub-block.
And we prove that FFN$'_{\text{SwiGLU}}(x\pi)=$FFN$_{\text{SwiGLU}}(x)\pi$.

\textbf{Sparse attention.}
GPT-3~\cite{gpt3} adopts sparse attention patterns in the Transformer layer~\cite{sparse-trn}.
Similarly, Longformer~\cite{longformer} was proposed to improve the memory efficiency for long context.
These alterations in attention are equivalent to modifying the masking $M$ matrix.
According to our proof, Theorem~\ref{thm:1} holds for any $M$ matrix, obviating the need for adjusting the parameter transformation.

\subsection{Multi-Modal Models}
\label{subsec:real-mm}

ViT~\cite{vit} divides an image into non-overlapping patches, and each patch is linearly embedded to create a sequence of token embeddings. 
These token embeddings serve as the input to the Transformer model.
Since \mysys does not rely on the preprocessing of the original data, it can seamlessly support ViT.
LLaVA~\cite{liu2023llava} takes both text and an image as inputs.
It employs a visual transformer to embed the image and subsequently connects them with the embeddings of the text input using a linear projection $x_v W$, where $x_v$ denotes the visual embedding.
To integrate LLaVA with STIP, we only need to transform the projection weight $W$ by $\pi_{v}^TW\pi_{t}$, where $\pi_{v}$ and $\pi_{t}$ denote the permutation matrices used for visual and textual transformer features, respectively.

\subsection{Mixture-of-Experts Models}
\label{subsec:real-moe}

Mixtral~\cite{jiang2024mixtral} integrates mixture-of-experts (MoE) into Transformer by constructing multiple feedforward sub-blocks (referred to as experts) in parallel, complemented by a router (or gating layer). 
The router determines the weights for the experts through $g(x)=xW_g$, where $W_g\in \mathbb{R}^{d\times e}$ and $e$ represents the number of experts. 
To support MoE, a simple transformation of $W_g$ suffices, accomplished by $\pi^T W_g$.

\subsection{Application Scope}
\label{subsec:real-scope}

For layers with learnable parameters, STIP requires them only to involve global matrix multiplication (e.g., linear, self-attention and feedforward) or token-wise aggregation (e.g., LayerNorm).
To give some counterexamples, STIP cannot be extended to convolutional and recurrent layers.

For layers without learnable parameters, STIP requires them to meet $f(x\pi)=f(x)\pi$ property, i.e., column-wise permutation equivariance.
For example, ReLU, GeLU, SoftMax, and Sigmoid activation layers.

\section{Evaluation}
\label{sec:exp}

We evaluate \mysys on various Transformer inference tasks using both real systems and public datasets.
Our key findings are outlined as follows:

\noindent \textbullet~\mysys demonstrates practical security concerning model parameters and user data and has no loss of accuracy ($\S$~\ref{subsec:sec-acc}).

\noindent \textbullet~\mysys achieves throughput levels comparable to unprotected full-cloud inference, outperforming state-of-the-art secure two-party protocols~\cite{hou2023ciphergpt, hao2022iron, chen2022x} by millions of times ($\S$~\ref{subsec:eff}).

\noindent \textbullet~\mysys exhibits efficiency in evaluating various microbenchmarks, including communication overhead, memory footprint, and effect of model split ($\S$~\ref{subsec:exp-micro}).

\subsection{Implementation}
\label{subsec:implement}

We implemented \mysys using PyTorch and HuggingFace~\cite{wolf2020huggingfaces} libraries.
Modern deep learning frameworks, including PyTorch, adopt a row-major memory layout.
To align with the memory layout, PyTorch performs matrix multiplication in the linear layer as $x W^T$ instead of $x W$.
Consequently, we transposed the previously introduced parameter transformation for implementation.
For permutation operations, we opted to generate a random permutation vector $\pi_{v}$ instead of a matrix $\pi_{m}$.
This vector is then used to index rows or columns, as in $W[:, \pi_{v}]$, which achieves the same effect as $W\pi_m$. 
The indexing-based approach is more computationally efficient than matrix multiplication.

\subsection{Experimental Setup}
\label{subsec:setup}

\begin{table}[t]
\centering
\caption{Summary of Testbeds and Transformer Models}
\label{tab:testbeds}
\begin{tabular}{@{}lcc@{}}
\toprule
Testbeds & Modality & Transformers \\ \midrule
\begin{tabular}[c]{@{}l@{}}Campus Security \\ Chatbot (CHAT)\end{tabular} & Text & GPT2/LLaMA2 \\ \midrule
\begin{tabular}[c]{@{}l@{}}Vehicle Cabin Scene\\ Understanding (CABIN)\end{tabular} & Image & ViT/LLaVA \\ \midrule
\begin{tabular}[c]{@{}l@{}}Simulator (SIMU)\end{tabular} & Text & BERT/Mixtral \\ \bottomrule
\end{tabular}
\end{table}

\textbf{Testbeds and Transformer models.}
We use three testbeds and six representative Transformer models for evaluation, see Tab.~\ref{tab:testbeds}.
(1) Campus Security Chatbot (CHAT).
To support natural language Q\&A for campus security, we developed a large language model-based chatbot.
We select pretrained LLaMA2-7b~\cite{touvron2023llama} to host this service at ANONYMOUS University.
To scale the evaluation, we also deployed GPT2-124m/355m/774m/1.5b~\cite{gpt2} and LLaMA2-13b/70b models\footnote{Note that the number after the connector - refers to the parameter amount.}.
(2) Vehicle Cabin Scene Understanding (CABIN).
At ANONYMOUS automotive company, we use LLaVA-13b~\cite{liu2023llava} to implement the cabin scene understanding function.
LLaVA model takes in-cabin video frames and a preset prompt as inputs to generate scene descriptions.
We also deployed ViT-86m/307m/632m models~\cite{vit} for comprehensive experiments.
(3) Simulator (SIMU).
To further evaluate \mysys on BERT series~\cite{bert} and Mixtral~\cite{jiang2024mixtral} models, we build a simulation testbed and test with randomly generated data.


\textbf{Baselines.}
For comparisons, we consider four approaches:
(1) Full-cloud.
Transformer models with original parameters are deployed on the cloud and the device sends raw data (plaintext) to the cloud for inference.
(2-4) Iron~\cite{hao2022iron}, THE-X~\cite{chen2022x}, and CipherGPT~\cite{hou2023ciphergpt}.
They propose secure two-party protocols for serving BERT series and GPT-2 models.

\textbf{Devices.}
For all cases, we use a server with 4 NVIDIA A100 GPUs as the model server.
In the CHAT testbed, we use a PC with 8-core Intel Core i7 CPUs as the user device.
In the CABIN testbed, we use an NVIDIA Orin development board as the user device.
And for SIMU, we use a MacBook Pro laptop with 4-core Intel Core i5 CPUs as the user device.

\subsection{Security and Accuracy Guarantee}
\label{subsec:sec-acc}

\begin{figure}[t]
    \centering
    \begin{subfigure}[b]{0.49\linewidth}
        \centering
        \includegraphics[width=\linewidth]{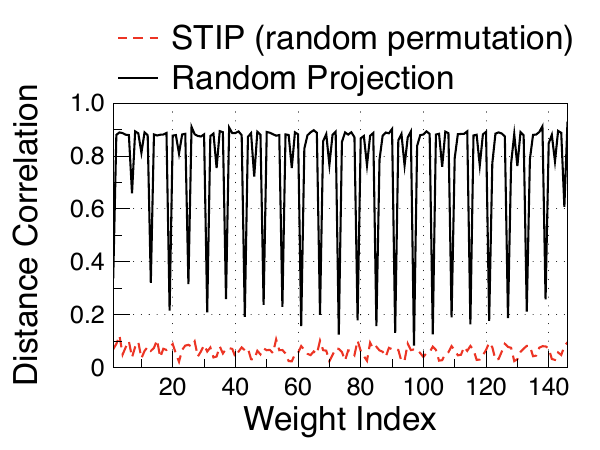}
        \caption{Parameters of GPT2}
        \label{fig:para-corr}
    \end{subfigure}
    \hfill
    \begin{subfigure}[b]{0.49\linewidth}
        \centering
        \includegraphics[width=\linewidth]{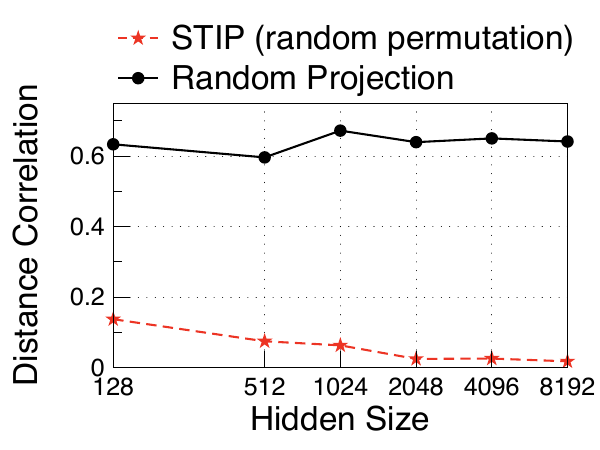}
        \caption{Embeddings}
        \label{fig:data-corr}
    \end{subfigure}
    \caption{Privacy leakage measurement: distance correlation.}
\end{figure}

First, we evaluate the previously proven security and computational equivalence through experiments.

\textbf{Distance correlation.}
In our privacy analysis, we employ distance correlation~\cite{distcorr} as the metric for assessing privacy leakage. 
As a baseline, we utilize random linear projection on both parameters and embeddings, referred to as Random Projection.
In Fig.~\ref{fig:para-corr}, we present the distance correlation between the original and transformed parameters of the GPT2-1.5b model. 
Notably, STIP demonstrates a significantly lower distance correlation compared to Random Projection. 
On average, Random Projection yields a distance correlation of 0.76, while STIP achieves a markedly lower value of 0.062.
To evaluate the security of on-device data, we apply transformations to embeddings with various hidden sizes ranging from 128 to 8192. 
The resulting distance correlations are depicted in Fig.~\ref{fig:data-corr}. 
In the case of Random Projection, the transformed data maintains a correlation higher than 0.6 on average. Conversely, the distance correlation of STIP diminishes with increasing hidden sizes, ranging from 0.14 to 0.017. 
This showcases the effectiveness of STIP in reducing privacy leakage associated with transformed data.
Our experimental findings affirm the low privacy leakage of permutation-based transformed data and parameters, providing validation for our bound analysis in Sec.~\ref{subsec:sec-ana}.

\textbf{On-cloud parameter unavailability.}
In addition to quantifying the correlation, we also assess the practical unavailability of on-cloud parameters by generating tokens using transformed parameters.
We use identical prompts and feed them to the GPT2-1.5b model with original and STIP transformed parameters.
Tab.~\ref{tab:parasecure} demonstrates the results.
With the prompt ``I'm a language model,'', tokens generated by on-cloud transformed parameters are completely meaningless, highlighting the practical unavailability of deployed parameters.
This observation emphasizes the effectiveness of \mysys in securing parameters from unauthorized use.

\begin{table}[t]
\centering
\caption{Unauthorized use of on-cloud transformed model generates meaningless tokens.}
\label{tab:parasecure}
\begin{tabular}{@{}ll@{}}
\toprule
\multicolumn{1}{l|}{Real Generation} & On-cloud Generation \\ \midrule
\multicolumn{2}{c}{Prompt: I'm a language model,} \\ \midrule
\multicolumn{1}{l|}{{\color[HTML]{000000} \begin{tabular}[c]{@{}l@{}}but what I do in that \\ role is to change \\ everything in our lives.\end{tabular}}} & 
    \begin{tabular}[c]{@{}l@{}}examines Blazers \\ consolationtechorate\\ applicationkiJanuary \\ PLANkikiorate Blazers \\ consolation Beyondki 
    \end{tabular} \\ \bottomrule
\end{tabular}
\end{table}

\textbf{No loss of accuracy.}
A key advantage of \mysys lies in its computational equivalence, ensuring that serving Transformer models with \mysys incurs no loss of accuracy. 
We assess this by examining two metrics: the sum of absolute differences in predictions and top-1 token classification accuracy.
We conducted tests on all six selected model series, ranging from 4 million to 70 billion parameters, using 10000 samples each.
As depicted in Table~\ref{tab:absdiff}, \mysys consistently achieves 100\% accuracy across all models. 
It's worth noting that the slight non-zero absolute difference is attributable to inherent floating-point operation errors rather than any loss of accuracy introduced by \mysys.

\begin{table*}[t]
\centering
\caption{STIP achieves no loss of Transformer inference accuracy.}
\label{tab:absdiff}
\begin{tabular}{@{}l|cccccc@{}}
\toprule
Model & GPT-2 & LLaMA2 & ViT & BERT & LLaVA & Mixtral \\
Paras. & 124m/355m/774m/1.5b & 7b/13b/70b & 86m/307m/632m & 4m/41m/110m/336m & 13b & 47b \\ \midrule
Sum of Abs. Diff. & 0.021/0.033/0.0478/0.051 & 0.009/0.012/0.012 & 3e-4/3e-4/3e-4 & 5e-3/8e-3/9e-3/9e-3 & 0.016 & 0.008 \\
Class Acc. & 100\% & 100\% & 100\% & 100\% & 100\% & 100\% \\ \bottomrule
\end{tabular}
\end{table*}

\subsection{Inference Efficiency}
\label{subsec:eff}

Next, we evaluate the inference efficiency of STIP.
The results are tested on the testbed devices associated with the model, and we will not make additional explanations.

\begin{figure}[t]
    \centering
    \begin{subfigure}[b]{0.49\linewidth}
        \centering
        \includegraphics[width=\linewidth]{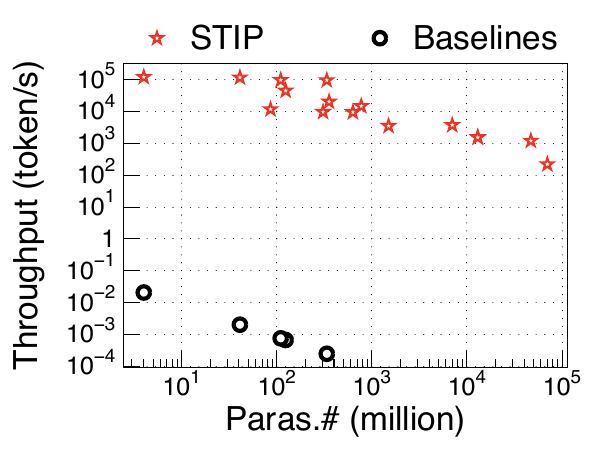}
        \caption{End-to-end Throughput}
    \end{subfigure}
    \hfill
    \begin{subfigure}[b]{0.49\linewidth}
        \centering
        \includegraphics[width=\linewidth]{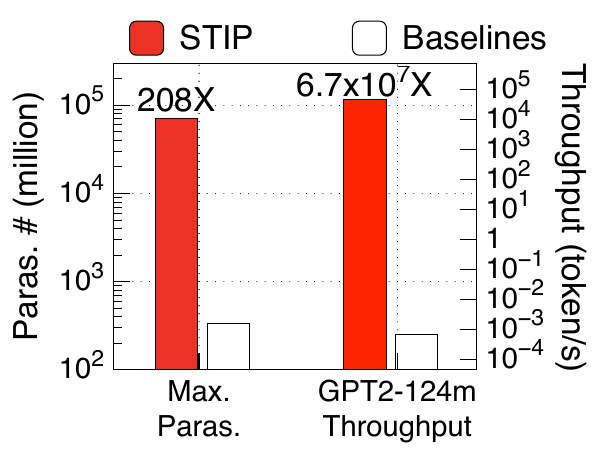}
        \caption{Improvement}
    \end{subfigure}
    \hfill
    \begin{subfigure}[b]{0.49\linewidth}
        \centering
        \includegraphics[width=\linewidth]{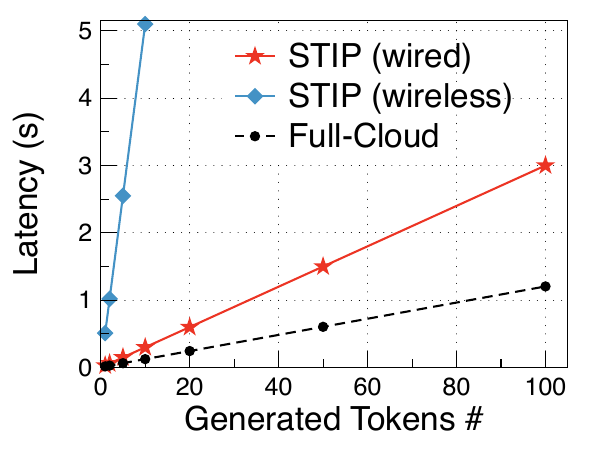}
        \caption{Autoregressive Generation}
    \end{subfigure}
    \hfill
    \begin{subfigure}[b]{0.49\linewidth}
        \centering
        \includegraphics[width=\linewidth]{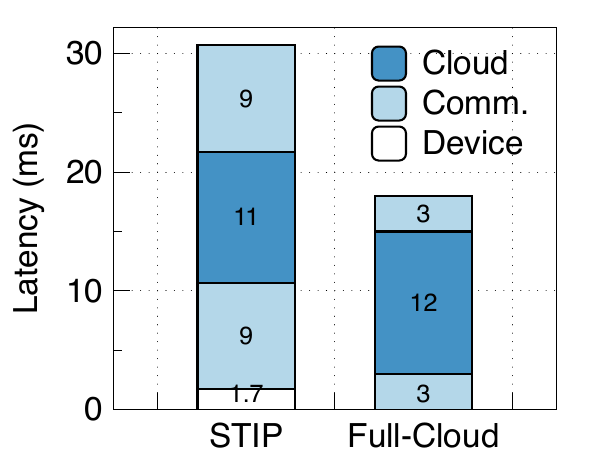}
        \caption{Latency Breakdown}
        \label{fig:lat-breakdown}
    \end{subfigure}
    \caption{Efficient serving Transformers with STIP.}
    \vspace{-0.2in}
    \label{fig:throughput}
\end{figure}

\textbf{End-to-end throughput and scalability with parameter size.}
We conducted tests to evaluate the end-to-end throughput of serving Transformer models with STIP. 
The batch size was set to 100, and the number of tokens per sample was set to 100.
As illustrated in Fig.~\ref{fig:throughput} (a), STIP demonstrates orders of magnitude higher throughput compared to baselines. 
Note that the baseline's throughput is inferred from reported results in its paper, as the secure protocols~\cite{hou2023ciphergpt, hao2022iron, chen2022x} lack open-source code. 
Additionally, we performed full-cloud inference tests, but the results were close to STIP, causing overlap of markers and, consequently, were omitted for clarity.
For the improvement of the maximal number of parameters, STIP reaches 70 billion compared to 336 million in the baselines, marking a 208-fold increase. 
For GPT2-124m throughput, CipherGPT reported 6.7e-4 token/s, whereas \mysys achieves 45,366 token/s, showcasing an improvement of 6.7 million times.
Fig.~\ref{fig:throughput} (b) summarizes the parameter scale and throughput improvements.

\textbf{Autoregressive generation.}
In addition to a single-round feedforward pass, we conducted tests on autoregressive generation with STIP, considering both wired and wireless network connections for STIP communication. 
The average communication latency for wired connections is approximately 10ms, while for wireless connections, it is around 250ms.
With a batch size of 1 and 128 input prompts, Fig.~\ref{fig:throughput} (c) presents the results for the LLaMA2-7b model. 
The latency for all serving approaches exhibits a linear increase with the number of generated tokens. 
The slopes for the result lines of Full-cloud, STIP wired, and STIP wireless are approximately 12, 30, and 510, respectively.
As discussed in Sec.~\ref{subsec:protocol}, the communication cost per generated token is inevitable to ensure output privacy protection. 
Considering the practical security that STIP introduces compared to unprotected full-cloud inference, the slightly higher latency (e.g., 2s more for 100 tokens) is deemed acceptable.

\textbf{Latency breakdown.}
To gain deeper insights into the overhead introduced by STIP, we conducted an analysis of latency breakdown, comparing it against full-cloud inference. 
Illustrated in Fig.~\ref{fig:lat-breakdown}, our evaluation reveals that STIP introduces an additional 1.7ms latency on the device while concurrently reducing on-cloud latency from 12ms to 11ms. 
A crucial factor contributing to the slower performance of STIP compared to full-cloud is the communication phase. 
This arises from the necessity of transmitting intermediate embeddings, a $\text{BATCH} \times n\times d$ tensor, which typically exceeds the size of plaintext words transmitted in full-cloud serving. 
While prior efforts~\cite{yao2020deepcod} have investigated techniques to compress intermediate activations and enhance communication efficiency in model splitting scenarios, it is noteworthy that our work imposes strict requirements for lossless accuracy, rendering these compression techniques beyond the current design scope. 
Integrating such compression methods with STIP is a promising direction for future research.

\subsection{Micro-Benchmarks}
\label{subsec:exp-micro}

\textbf{Device-cloud communication traffic.}
The communication traffic induced by STIP is influenced by three factors: the number of input tokens, hidden size, and output vocabulary size. 
To illustrate, considering the GPT2-124m model, a single-round inference operation causes 5.8 MiB and 7.5 MiB of traffic for input embedding and output activations, respectively. 
As depicted in Fig.~\ref{fig:challenge-overhead}, the communication traffic incurred by STIP is markedly lower compared to CipherGPT, 95,151 MiB. 
This substantial reduction in traffic highlights STIP's ability to achieve security at a modest cost. 

\begin{figure}[t]
    \centering
    \begin{subfigure}[b]{0.8\linewidth}
        \centering
        \includegraphics[width=\linewidth]{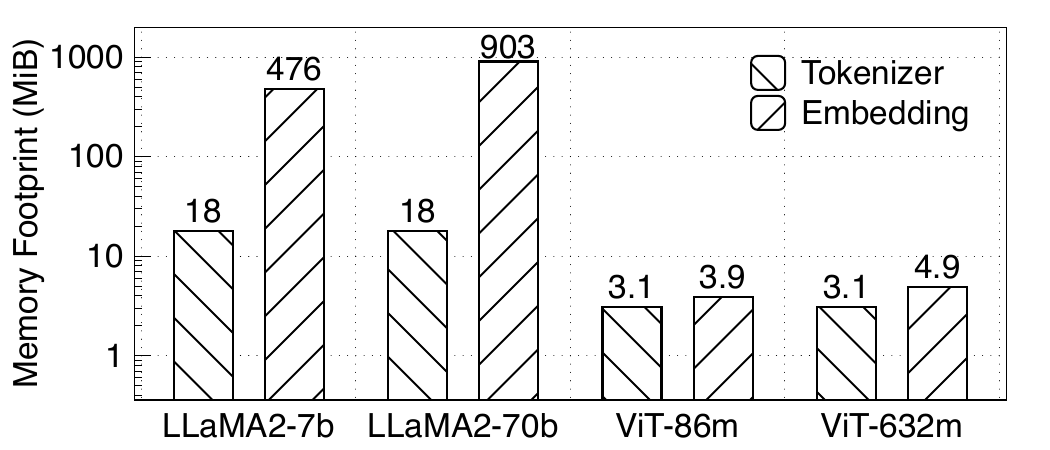}
    \end{subfigure}
    \caption{On-device memory usage.}
    \vspace{-0.1in}
    \label{fig:devicemem}
\end{figure}

\textbf{On-device memory footprint.}
In light of the diverse range of devices that may be employed for Transformer-based services, we assess the on-device memory footprint. 
The results, depicted in Fig.~\ref{fig:devicemem}, showcase the memory requirements.
For the tokenizer component, LLaMA2 and ViT models exhibit memory footprints of 18 MiB and 3.1 MiB, respectively. 
In the case of the embedding part, the memory allocation depends on the hidden size parameter. 
LLaMA2-70b, utilizing a large hidden size of 8192, incurs a memory cost of 903 MiB. 
In contrast, the ViT models exhibit more modest memory requirements, ranging from 3.9 MiB to 4.9 MiB.
This implies that the on-device memory demands of STIP, even for models with substantial hidden sizes, remain feasible for contemporary end devices.

\begin{figure}[t]
    \centering
    \begin{subfigure}[b]{0.8\linewidth}
        \centering
        \includegraphics[width=\linewidth]{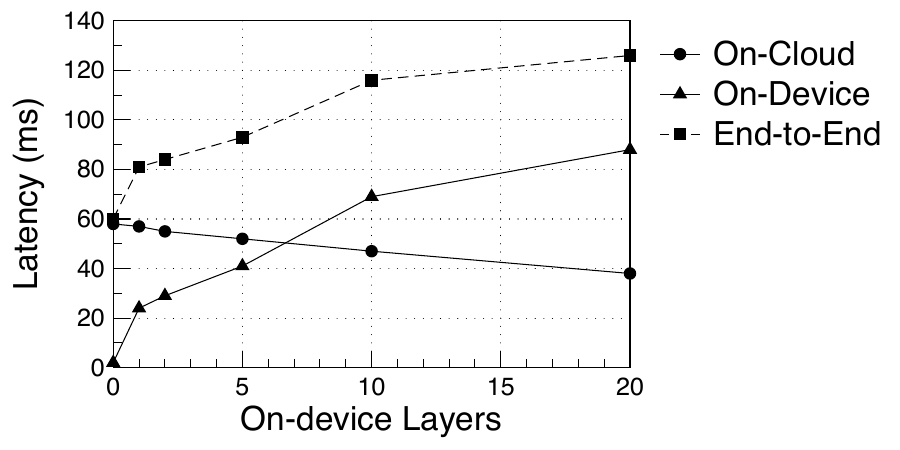}
    \end{subfigure}
    \caption{Latency with different model splits.}
    \vspace{-0.2in}
    \label{fig:splitlat}
\end{figure}

\textbf{Effect of model split.}
We vary the number of on-device Transformer layers from 0 to 20 and analyze the corresponding impact on inference latency.
As depicted in Fig.\ref{fig:splitlat}, the latency of end-to-end inference rises proportionally with an increasing number of on-device layers. 
This latency increase is attributed to the relatively lower computing power of devices compared to the cloud. 
As discussed in Sec.~\ref{subsec:sec-ana}, deploying more layers on the device not only results in higher latency but also exposes additional parameters to the user, thereby introducing privacy risks.
In light of these considerations, our analysis indicates that deploying only the embedding module on the device represents the optimal choice. 
This configuration minimizes latency while mitigating the potential privacy risks of exposing more layers to the user.

\section{Discussion}
\label{sec:discuss}

\textbf{Cloud Providers with Internal Model Development.}
A notable limitation of \mysys stems from its three-party setting. 
While this setting proves suitable for a range of applications, as detailed in Sec.~\ref{subsec:threep}, it may not be well-suited when cloud providers are also involved in model development.
Consider the scenario where cloud providers, like Google, maintain their own cloud platform along with large Transformer models~\cite{gemini, bard}. 
Despite the possibility of implementing access control within different departments of the same company to avoid collusion, there remains a level of skepticism regarding external users. 
The inherent challenge lies in establishing a trustworthy environment, especially when external users cannot guarantee the separation of interests and potential conflicts of interest within the cloud provider's internal model development and hosting processes. 

\noindent \textbf{Training support extension.}
STIP is designed to address the forward pass, and model training introduces the complexity of backward gradient propagation. 
We envision that the principles underlying STIP could be promisingly extended to support privacy-preserving training.
However, this extension entails studying the communication overhead associated with gradients and the additional privacy leakage risks introduced during gradient propagation~\cite{wang2022protect}. 
Future ongoing exploration is needed to fully realize the potential of STIP in the context of privacy-preserving Transformer training.

\section{Related Work}

\noindent \textbf{Neural network split.}
The practice of splitting neural network layers and distributing them between the device and server has been explored as a means to protect raw on-device data while preserving efficiency~\cite{dc-transformer, thapa2022splitfed, zeng2020coedge, pham2023binarizing, pasquini2021unleashing}. 
However, despite this split, the potential for reverse-engineering sensitive information from intermediate activations~\cite{abuadbba2020can}, such as text embeddings~\cite{pan2020privacy}, remains a concern.
STIP builds upon the concept of model split and goes a step further by incorporating random permutation, offering theoretically enhanced security measures. 

\noindent \textbf{Secure Transformer inference.}
In the context of a two-party setting, prior efforts~\cite{hou2023ciphergpt, hao2022iron, chen2022x} have explored the combination of homomorphic encryption and multi-party computation techniques to devise secure protocols for Transformer inference. 
These approaches customize and optimize computation protocols for specific layers within Transformer models, such as non-linear activation and layer normalization.
In contrast to these two-party systems, STIP introduces a novel three-party setting and employs a semi-symmetrical permutation scheme to enhance security. 

\section{Conclusion}
In this paper, we studied security concerns in Transformer inference.
We proposed a three-party threat model and presented the design of STIP, a secure transformer inference protocol based on our semi-symmetrical permutation scheme.
Theoretical analysis and experiments in real systems evaluated the security, computational equivalence, and practical efficiency of STIP.

\bibliographystyle{ACM-Reference-Format}
\bibliography{reference}

\newpage
\appendix

\noindent \textbf{Appendices are supporting material that has not been peer-reviewed.}
\section{Proofs}
\label{appendix:proof}

\noindent \textbf{Proof of Theorem~\ref{thm:1}:}

\begin{equation*}
    F_{\theta'}(x\pi)\pi_c^T = F_\theta(x).
\end{equation*}

\begin{proof}
    First, since the calculation of non-linear activation is element-wise, they are permutation equivalent, i.e., ReLU$(x\pi)$=ReLU$(x)\pi$ and SoftMax$(x\pi)$=SoftMax$(x)\pi$.
    
    \noindent Next, we prove that:
    
    LayerNorm$(x\pi;\gamma\pi, \beta\pi)=$LayerNorm$(x;\gamma, \beta)\pi$.
    
    \noindent The LayerNorm function is defined for $x \in \mathbb{R}^{n\times d}$ by
    \begin{align*}
        &\text{LayerNorm}(x; \gamma, \beta) = \gamma \circ \frac{x-\mu_x}{\sigma_x} + \beta, &\gamma,\beta \in \mathbb{R}^{d},
    \end{align*}
    where $\circ$ denotes the Hadamard (element-wise) product operator.
    Since $\mu_x$ and $\sigma_x$ are computed by rows, $\mu_{x\pi}=\mu_x$ and $\sigma_{x\pi} = \sigma_x$.
    Therefore, 
    \begin{align*}
        \text{LayerNorm}(x\pi; \gamma\pi, \beta\pi) &= \gamma\pi \circ \frac{x\pi-\mu_x}{\sigma_x} + \beta \pi\\ 
        &= \left(\gamma \circ \frac{x-\mu_x}{\sigma_x} + \beta\right) \pi\\ 
        &= \text{LayerNorm}(x;\gamma, \beta)\pi.
    \end{align*}
    Then, since $\forall \pi, \pi \pi^T=I$:
    \begin{align*}
        Q' &= x\pi \pi^T W_q \pi_{i,1} = x W_q \pi_{i,1} = Q \pi_{i,1},\\
        K' &= x\pi \pi^T W_k \pi_{i,1} = x W_k \pi_{i,1} = K \pi_{i,1},\\
        V' &= x\pi \pi^T W_v \pi_{i,2} = x W_v \pi_{i,2} = V \pi_{i,2},\\
        u' &= \text{SoftMax}\left( \frac{Q'K'^T}{\sqrt{k}} + M\right)V' \pi_{i,2}^T W_o\pi\\ 
        &= \text{SoftMax}\left( \frac{Q\pi_{i,1} \pi_{i,1}^T K^T}{\sqrt{k}} + M\right)V \pi_{i,2} \pi_{i,2}^T W_o\pi\\
        &= \text{SoftMax}\left( \frac{QK^T}{\sqrt{k}} + M\right)V W_o\pi = u \pi,\\
        v' &= \text{LayerNorm}(u' + x\pi; \gamma_1', \beta_1')\\
        &=\text{LayerNorm}(u\pi + x\pi; \gamma_1\pi, \beta_1\pi)\\
        &=\text{LayerNorm}((u + x)\pi; \gamma_1\pi, \beta_1\pi)=v\pi,\\
        z' &= \text{ReLU}(v' \pi W_1')W_2'=\text{ReLU}(v\pi \pi^T W_1 \pi_{i,3})\pi_{i,3}^T W_2 \pi\\
        &=\text{ReLU}(vW_1)W_2 \pi=z\pi,\\
        y' &= \text{LayerNorm}(z' + v'; \gamma_2', \beta_2')\\
        &=\text{LayerNorm}(z\pi + v\pi; \gamma_2\pi, \beta_2\pi)\\
        &=\text{LayerNorm}((z + v)\pi; \gamma_2\pi, \beta_2\pi)=y\pi,\\
        o' &= \text{SoftMax}(y'W_c') = \text{SoftMax}(y\pi \pi^T W_c \pi_c) = o\pi_c.
    \end{align*}
    Therefore, $F_\theta'(x\pi)\pi_c^T=o'\pi_c^T=o\pi_c\pi_c^T=o=F_\theta(x)$.
\end{proof}

\noindent \textbf{Proof of Pre-LayerNorm.}

\begin{proof}
    From the proof of Theorem.1, we can see the permutation equivalence property holds for the self-attention sub-block, i.e., $\text{Attn}(x\pi)=\text{Attn}(x)\pi$.
    So
    \begin{align*}
        v' &= \text{Attn}(\text{LayerNorm}'(x\pi))+x\pi\\
        &= \text{Attn}(\text{LayerNorm}(x)\pi)+x\pi\\
        &= \text{Attn}(\text{LayerNorm}(x))\pi+x\pi\\
        &= (\text{Attn}(\text{LayerNorm}(x))+x)\pi = v\pi,\\
        y' &= \text{ReLU}(\text{LayerNorm}'(v')W_1')W_2'+v'\\
        &= \text{ReLU}(\text{LayerNorm}'(v\pi)\pi^T W_1 \pi_{i,3}) \pi_{i,3} ^T W_2 \pi +v\pi\\
        &= (\text{ReLU}(\text{LayerNorm}(v)W_1)W_2+v)\pi = y\pi,
    \end{align*}
    where LayerNorm$'$ denotes layer normalization with transformed parameters.
    
    \noindent Therefore, $F_\theta'(x\pi)\pi_c^T=F_\theta(x)$ still holds.
\end{proof}

\noindent \textbf{Proof of RMSNorm.}

\begin{proof}
    The RMSNorm function is defined for $x\in \mathbb{R}^{n\times d}$ by
    \begin{align*}
        &\text{RMSNorm}(x; \gamma) = \gamma \circ \frac{x}{\sqrt{\frac{1}{n}\sum_i x_i^2}}, &\gamma\in \mathbb{R}^{d},
    \end{align*}
    where $\circ$ denotes the Hadamard (element-wise) product operator.
    Since $\sum_i x_i^2$ is computed by rows, $\sum_i (x\pi)_i^2 = \sum_i x_i^2$.
    Therefore, 
    \begin{align*}
        \text{RMSNorm}(x\pi; \gamma\pi) &= \gamma\pi \circ \frac{x\pi}{\sqrt{\frac{1}{n}\sum_i (x\pi)_i^2}}\\ 
        &= \left(\gamma \circ \frac{x}{\sqrt{\frac{1}{n}\sum_i x_i^2}}\right) \pi\\ 
        &= \text{RMSNorm}(x;\gamma)\pi.
    \end{align*}
\end{proof}

\noindent \textbf{Proof of SwiGLU feedforward.}

\begin{proof}
    By definition,
    \begin{align*}
    &\text{FFN}'_{\text{SwiGLU}}(x\pi)= (x\pi W_1' \text{Sigmoid}(x\pi W_1') x\pi W_3) W_2'\\
    &=(x\pi \pi^TW_1 \text{Sigmoid}(x\pi \pi^TW_1) x\pi \pi^TW_3 \pi_{i,3}) \pi_{i,3}^T W_2\pi\\ 
    &= (xW_1\text{Sigmoid}(xW_1)xW_3)W_2\pi\\
    &= \text{FFN}_{\text{SwiGLU}}(x)\pi.
\end{align*}
\end{proof}

\end{document}